\documentclass{article}
\usepackage[T1]{fontenc} 
\usepackage[utf8]{inputenc} 
\usepackage{ismir,amsmath,cite,url}
\usepackage{graphicx}
\usepackage{algorithm}
\usepackage[noend]{algpseudocode}
\usepackage{caption}
\usepackage{booktabs}
\usepackage{array}

\usepackage{makecell}

\usepackage{color}

\usepackage[firstpage]{draftwatermark}
\definecolor{lightgray}{rgb}{0.9,0.9,0.9}
\definecolor{darkgray}{rgb}{0.4,0.4,0.4}
\SetWatermarkFontSize{12pt}
\SetWatermarkScale{1.1}
\SetWatermarkAngle{90}
\SetWatermarkHorCenter{202mm}
\SetWatermarkVerCenter{170mm}
\SetWatermarkColor{darkgray}
\SetWatermarkText{Late-Breaking / Demo Session Extended Abstract, ISMIR 2024 Conference}

\def\lbd{}	

\title{
Local deployment of large-scale music AI models on commodity hardware
}

\multauthor
{Xun Zhou \hspace{1cm} Charlie Ruan \hspace{1cm} Zihe Zhao \hspace{1cm} Tianqi Chen \hspace{1cm} Chris Donahue} {
Carnegie Mellon University
}

\def\authorname{Zhou, Ruan, Zhao, Chen, Donahue}

\usepackage[bookmarks=false,pdfauthor={\authorname},pdfsubject={\papersubject},hidelinks]{hyperref}
\usepackage{cleveref}

\begin{document}

\maketitle
%

\hspace{-4mm} \url{https://rickzx.github.io/inf-music}

\begin{abstract}
We present the \emph{MIDInfinite}, a web application capable of generating symbolic music using a large-scale generative AI model 
locally on commodity hardware. 
Creating this demo involved porting the Anticipatory Music Transformer, a large language model (LLM) pre-trained on the Lakh MIDI dataset, to the Machine Learning Compilation (MLC) framework. 
Once the model is ported, MLC facilitates inference on a variety of runtimes including C++, mobile, and the browser. 
We envision that MLC has the potential to bridge the gap between the landscape of increasingly capable music AI models and technology more familiar to music software developers. 
As a proof of concept, 
we build a web application that allows users to generate endless streams of multi-instrumental MIDI in the browser, either from scratch or conditioned on a prompt. 
On commodity hardware (an M3 Macbook Pro), our demo can generate $51$ notes per second, which is faster than real-time playback for $72.9$\% of generations, and increases to $86.3$\% with $2$ seconds of upfront buffering.
\end{abstract}
\vspace{-2mm}
\section{Introduction}\label{sec:introduction}


Large-scale generative AI models for music are increasingly capable and have the potential to transform both music technology and music creation workflows. 
However, models are currently tightly coupled to a software ecosystem (e.g.,~Python, CLIs) distinct from that of music technology (e.g.,~C++, VST plugins), 
inhibiting music software developers from exploring this potential. 
Moreover, models may even require specialized hardware (GPUs) for deployment, 
a prerequisite that may be unrealistic for many musicians. 
Cloud inference via APIs could help to bridge this gap~\cite{garcia2023harp}, 
but music creativity benefits from low latency, high reliability, and increased privacy, all of which are difficult to guarantee with APIs. 

\begin{figure*}[tp]
 \centerline{
 \includegraphics[alt={Deploy symbolic music generation model on-device with MLC},width=2\columnwidth]{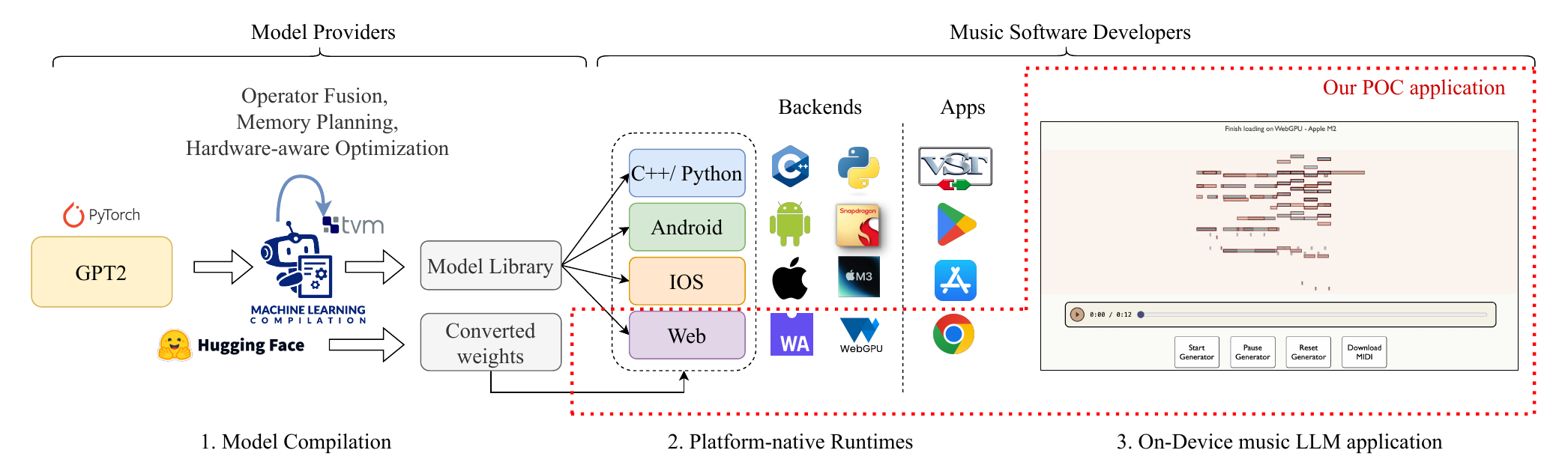}}
 \caption{Proposed workflow for bridging the gap between large model and music software ecosystems, which involves contributions from both model providers (porting models to MLC) and music software developers (building applications on resultant runtimes). Here we build a web demo of a state-of-the-art symbolic music generation model as a proof-of-concept.}
 \label{fig:workflow}
\end{figure*}

In this work, we propose a realistic workflow for deploying large-scale generative AI models on device in the music technology ecosystem (\Cref{fig:workflow}). 
This workflow involves decentralized contributions from two key groups of stakeholders: model providers and music software developers. 
Model providers are tasked with porting their models to the MLC-LLM framework~\cite{mlc-llm, tensorir, metaschedule, tvm, relax}, which facilitates both model compilation and provides a variety of runtimes for deployment. 
Accordingly, MLC-LLM helps overcome two key obstacles associated with the music technology ecosystem: 
model compilation allows models to run faster on commodity hardware, and 
runtimes bridge the gap to technology stacks more familiar to music software developers (C++, web, mobile). 
Music software developers are then tasked with creating rich interactive music applications that incorporate these generative capabilities. 
Finally, musicians can then explore these applications on commodity hardware such as laptops and mobile devices in familiar environments like DAWs. 
Related work~\cite{garcia2021deep} allows model providers to deploy to a specific music application---while our goal is to enable music software developers to explore deployment across a multitude of music applications. 


We present a proof-of-concept workflow, porting a state-of-the-art symbolic music generation model to MLC-LLM (assuming the role of model providers), 
and developing a web application using the WebGPU runtime (assuming the role of music software developers). 
Our application is capable of generating an endless stream of multi-instrumental MIDI in the browser on commodity hardware, either from scratch or conditioned on existing MIDI.

\begin{figure*}[ht]
    \centering
    \begin{minipage}{0.48\linewidth}
        \centering
        \scriptsize
        \begin{tabular}{@{}lccc@{}}
        \toprule
            \textbf{Config.}
            & \makecell{\textbf{Throughput} \\ tok/s}
            & \makecell{\textbf{Streamable} \\ (\%)}
            & \makecell{\textbf{Streamable} \\ (\%), w/ 2s buffer} \\
        \midrule
        \textbf{Small 128M} & & & \\
        ~~M2 w/ PyTorch   & 44  & 3.9 & 13.7 \\
        ~~M2 w/ MLC     & 84   & 24.1 & 47.0 \\
        ~~M3 Max w/ Pytorch     & 92  & 30.8  & 54.3 \\ 
        ~~M3 Max w/ MLC     & 155  & 72.9  & 86.3 \\ \bottomrule
        \textbf{Medium 360M} & & & \\
        ~~M2 w/ PyTorch         &   14       & 0.2  & 1.9 \\
        ~~M2 w/ MLC     & 38   & 2.3 & 10.2 \\
        ~~M3 Max w/ Pytorch     & 31   & 1.2 & 6.7 \\
        ~~M3 Max w/ MLC     & 90  & 28.7  & 52.6  \\ \midrule
        \end{tabular}
        \captionof{table}{Profiling different models, runtimes, and commodity chips. Streamable is \% of time where time in generation stream exceeds time in playback stream, with and without an initial $2$s playback buffer.}
        \label{tab:model_performance}
    \end{minipage}%
    \hfill
    \begin{minipage}{0.48\linewidth}
        \centering
        \includegraphics[width=\linewidth]{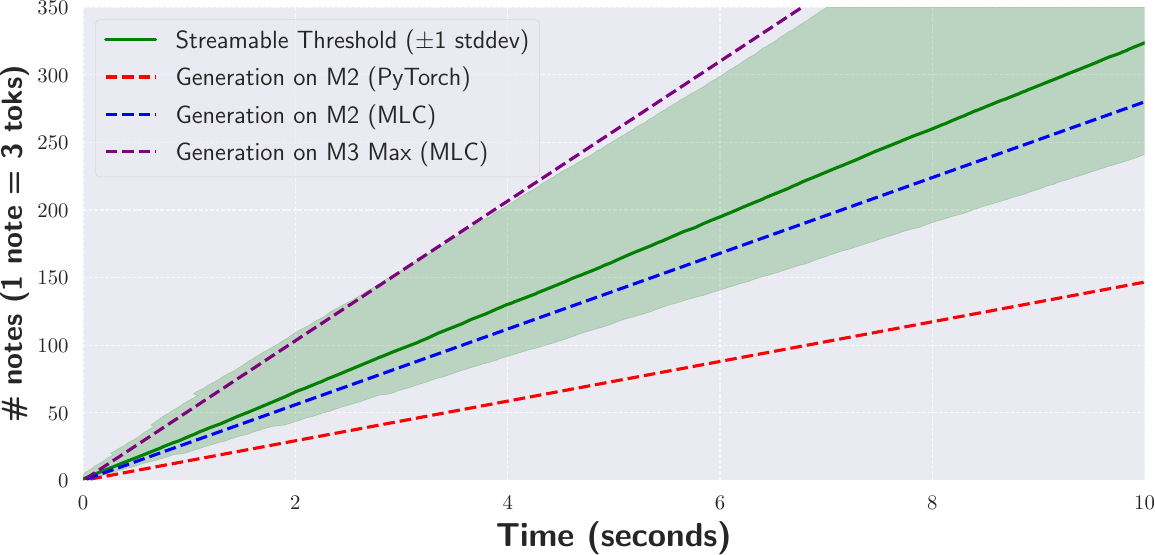}
        \caption{Streaming performance visualization for small model. To stream, chips (dashed lines) must exceed the tok/s of music playback (solid line), which varies based on generated note density (shaded $\pm$1 stdev.).
        }
        \label{fig:token_generation}
    \end{minipage}
\vspace{-4mm}
\end{figure*}

\vspace{-4mm}
\section{Porting a music AI model to MLC-LLM}

We select the Anticipatory Music Transformer~\cite{thickstun2023anticipatory}, 
a state-of-the-art symbolic music generation model pre-trained on the Lakh MIDI dataset~\cite{raffel2016learning}, 
as our target model for deployment. 
This model is available in a small~(128M) and medium~(360M) variant, reusing the architecture and hyperparameters of the GPT-2~\cite{radford2019language} small and medium configurations respectively. 
Each MIDI note is encoded as a triplet of tokens representing
(i)~onset time,
(ii)~note duration, and 
(iii)~instrument and pitch. 
The model was trained on sequences of $1024$ tokens: one start-of-sequence token followed by $1023$ tokens ($341$ MIDI note triplets). 

To port this model to MLC, we define the GPT-2 model architecture in a computational graph in TVM's Relax frontend, which is translated to Relax IR for optimization and compilation \cite{relax}.
We also convert the original model weights into an MLC-compatible format. 
We release our ported model to Github 
and converted weights to Huggingface
(links embedded in footer of demo page). 
Once the model is compiled, it can be deployed across various hardware platforms using MLC’s platform-native runtime. For instance, the runtime leverages \textit{WebGPU} for browser-based execution and \textit{Metal} for Apple devices to achieve efficient performance across platforms (Table~\ref{fig:workflow}).

\vspace{-4mm}
\section{Building MIDInfinite}

As a proof-of-concept showcasing local deployment, 
we build a web application that allows users to generate an infinite stream of MIDI from the Anticipatory Music Transformer. 
We build on the recent WebLLM~\cite{web-llm} platform to deploy MLC-compiled LLMs in the browser.

\textbf{Infinite chunk-wise generation.} 
The Anticipatory Music Transformer was not designed for streaming application---it only natively supports generation of up to $341$ notes at a time. 
To adapt this model to the streaming setting, we generate music in $170$ note chunks. 
For the first chunk, we simply generate $170$ notes starting from scratch, concatenating the output to the stream. 
For subsequent chunks, we take the most recent $170$ notes from the stream, relativize their offsets to start at $0$ seconds, and generate $170$ subsequent notes. 


\textbf{Supporting grammars in WebLLM.} 
We extend WebLLM to support the enforcement of context-free grammars during inference. 
The Anticipatory Music Transformer has a unique triplet grammar that the model is not guaranteed to conform to. 
However, we can enforce this grammar by constraining generation to certain subsets of the vocabulary at particular token indices. 
To accommodate this, we extend WebLLM with a \texttt{LogitProcessor} abstraction, allowing developers to enforce grammars by freely modifying model output logits.



\textbf{Enforcing ensemble density.} 
We also add the ability for users to specify particular ensembles (collections of instruments) to generate from. 
While the \texttt{LogitProcessor} can constrain generation to particular instruments, 
there is no guarantee that the model will generate notes for \emph{all} of those instruments. 
Therefore, we propose a simple controllable heuristic to encourage the model to use all of the instruments in the ensemble. 
Our heuristic works by 
biasing the note logits 
for each instrument in the ensemble proportional to the amount of time since a note was last generated for that instrument. 
By adjusting this additive factor, users can gently or forcefully encourage the model to use all the instruments. 




\vspace{-4mm}
\section{Performance analysis}

Table~\ref{tab:model_performance} presents token generation throughput (in tok/s) for two models (small and medium) on different chips and backends. 
Additionally, we report the \% of generations that are streamable. 
Specifically, we first estimate the empirical distribution of tokens per second in generated MIDI (varies depending on note density) by averaging across $500$ generations, 
and report the proportion of time where generated tok/s (roughly constant) would exceed playback tok/s (variable)---see Figure~\ref{fig:token_generation} for a visualization. 
We also report this number with $2$ seconds of buffering, simulating a user experience of a brief pause before playback.









\bibliography{ISMIR2024_lbd}

\begin{thebibliography}{10}
\providecommand{\url}[1]{#1}
\csname url@samestyle\endcsname
\providecommand{\newblock}{\relax}
\providecommand{\bibinfo}[2]{#2}
\providecommand{\BIBentrySTDinterwordspacing}{\spaceskip=0pt\relax}
\providecommand{\BIBentryALTinterwordstretchfactor}{4}
\providecommand{\BIBentryALTinterwordspacing}{\spaceskip=\fontdimen2\font plus
\BIBentryALTinterwordstretchfactor\fontdimen3\font minus \fontdimen4\font\relax}
\providecommand{\BIBforeignlanguage}[2]{{%
\expandafter\ifx\csname l@#1\endcsname\relax
\typeout{** WARNING: IEEEtran.bst: No hyphenation pattern has been}%
\typeout{** loaded for the language `#1'. Using the pattern for}%
\typeout{** the default language instead.}%
\else
\language=\csname l@#1\endcsname
\fi
#2}}
\providecommand{\BIBdecl}{\relax}
\BIBdecl

\bibitem{garcia2023harp}
H.~F. Garcia, P.~O’Reilly, A.~Aguilar, B.~Pardo, C.~Benetatos, and Z.~Duan, ``{HARP}: Bringing deep learning to the daw with hosted, asynchronous, remote processing,'' \emph{Machine Learning for Creativity and Design, NeurIPS}, 2023.

\bibitem{mlc-llm}
\BIBentryALTinterwordspacing
M.~team, ``{MLC-LLM},'' 2023. [Online]. Available: \url{https://github.com/mlc-ai/mlc-llm}
\BIBentrySTDinterwordspacing

\bibitem{tensorir}
\BIBentryALTinterwordspacing
S.~Feng, B.~Hou, H.~Jin, W.~Lin, J.~Shao, R.~Lai, Z.~Ye, L.~Zheng, C.~H. Yu, Y.~Yu, and T.~Chen, ``Tensorir: An abstraction for automatic tensorized program optimization,'' in \emph{Proceedings of the 28th ACM International Conference on Architectural Support for Programming Languages and Operating Systems, Volume 2}, ser. ASPLOS 2023.\hskip 1em plus 0.5em minus 0.4em\relax New York, NY, USA: Association for Computing Machinery, 2023, p. 804–817. [Online]. Available: \url{https://doi.org/10.1145/3575693.3576933}
\BIBentrySTDinterwordspacing

\bibitem{metaschedule}
\BIBentryALTinterwordspacing
J.~Shao, X.~Zhou, S.~Feng, B.~Hou, R.~Lai, H.~Jin, W.~Lin, M.~Masuda, C.~H. Yu, and T.~Chen, ``Tensor program optimization with probabilistic programs,'' in \emph{Advances in Neural Information Processing Systems}, S.~Koyejo, S.~Mohamed, A.~Agarwal, D.~Belgrave, K.~Cho, and A.~Oh, Eds., vol.~35.\hskip 1em plus 0.5em minus 0.4em\relax Curran Associates, Inc., 2022, pp. 35\,783--35\,796. [Online]. Available: \url{https://proceedings.neurips.cc/paper_files/paper/2022/file/e894eafae43e68b4c8dfdacf742bcbf3-Paper-Conference.pdf}
\BIBentrySTDinterwordspacing

\bibitem{tvm}
\BIBentryALTinterwordspacing
T.~Chen, T.~Moreau, Z.~Jiang, L.~Zheng, E.~Yan, H.~Shen, M.~Cowan, L.~Wang, Y.~Hu, L.~Ceze, C.~Guestrin, and A.~Krishnamurthy, ``{TVM}: An automated {End-to-End} optimizing compiler for deep learning,'' in \emph{13th USENIX Symposium on Operating Systems Design and Implementation (OSDI 18)}.\hskip 1em plus 0.5em minus 0.4em\relax Carlsbad, CA: USENIX Association, Oct. 2018, pp. 578--594. [Online]. Available: \url{https://www.usenix.org/conference/osdi18/presentation/chen}
\BIBentrySTDinterwordspacing

\bibitem{relax}
\BIBentryALTinterwordspacing
R.~Lai, J.~Shao, S.~Feng, S.~S. Lyubomirsky, B.~Hou, W.~Lin, Z.~Ye, H.~Jin, Y.~Jin, J.~Liu, L.~Jin, Y.~Cai, Z.~Jiang, Y.~Wu, S.~Park, P.~Srivastava, J.~G. Roesch, T.~C. Mowry, and T.~Chen, ``Relax: Composable abstractions for end-to-end dynamic machine learning,'' 2023. [Online]. Available: \url{https://arxiv.org/abs/2311.02103}
\BIBentrySTDinterwordspacing

\bibitem{garcia2021deep}
H.~F. Garcia, A.~Aguilar, E.~Manilow, D.~Vedenko, and B.~Pardo, ``Deep learning tools for {Audacity}: Helping researchers expand the artist's toolkit,'' \emph{arXiv preprint arXiv:2110.13323}, 2021.

\bibitem{thickstun2023anticipatory}
J.~Thickstun, D.~Hall, C.~Donahue, and P.~Liang, ``Anticipatory music transformer,'' \emph{arXiv preprint arXiv:2306.08620}, 2023.

\bibitem{raffel2016learning}
C.~Raffel, \emph{Learning-based methods for comparing sequences, with applications to audio-to-midi alignment and matching}.\hskip 1em plus 0.5em minus 0.4em\relax Columbia University, 2016.

\bibitem{radford2019language}
A.~Radford, J.~Wu, R.~Child, D.~Luan, D.~Amodei, I.~Sutskever \emph{et~al.}, ``Language models are unsupervised multitask learners,'' \emph{OpenAI blog}, 2019.

\bibitem{web-llm}
\BIBentryALTinterwordspacing
M.~team, ``{WebLLM},'' 2023. [Online]. Available: \url{https://github.com/mlc-ai/web-llm}
\BIBentrySTDinterwordspacing

\end{thebibliography}

%
%
%
%
%

\end{document}